\documentclass[aps,prd,twocolumn,superscriptaddress]{revtex4}
\usepackage{epsfig,epsf}
\usepackage{amsmath}
\usepackage{amsthm}
\usepackage{amsfonts}
\usepackage{amssymb}
\usepackage{dsfont}
\usepackage{multirow}
\usepackage{appendix}
\usepackage{slashed}
\usepackage[active]{srcltx}
\usepackage{psfrag}

\begin{document}

\title{Hidden charm-bottom structures $bc\overline{b}\overline{c}$:
Axial-vector case}
\date{\today}
\author{S.~S.~Agaev}
\affiliation{Institute for Physical Problems, Baku State University, Az--1148 Baku,
Azerbaijan}
\author{K.~Azizi}
\affiliation{Department of Physics, University of Tehran, North Karegar Avenue, Tehran
14395-547, Iran}
\affiliation{Department of Physics, Do\v{g}u\c{s} University, Dudullu-\"{U}mraniye, 34775
Istanbul, T\"{u}rkiye}
\author{H.~Sundu}
\affiliation{Department of Physics Engineering, Istanbul Medeniyet University, 34700
Istanbul, T\"{u}rkiye}

\begin{abstract}
Mass and width of a hidden charm-bottom axial-vector structure $T$
containing $bc \overline{b}\overline{c}$ quarks are calculated in QCD sum
rule framework. It is treated as a diquark-antidiquark state built of scalar
diquark and axial-vector antidiquark components. The mass of $T$ is computed
using the two-point sum rule method. The width of this particle is evaluated
by considering eight decay modes: The decays to $\eta _{b}J/\psi $, $\eta
_{c}\Upsilon (1S)$, $B_{c}^{-}B_{c}^{\ast +}$, and $B_{c}^{+}B_{c}^{\ast -}$
are dissociation processes, in which all initial quarks are distributed
between the final-state particles. The decays to $DD$ and $BB$ mesons with
appropriate charges and spin-parities are channels generated due to the
annihilations of $b\overline{b}$ and $c\overline{c}$ quarks from $T$.
Partial widths for all of these processes are obtained by employing the
three-point sum rule approach necessary to find the strong couplings at
relevant tetraquark-meson-meson vertices. Our results for the mass $%
m=(12715\pm 90)~\mathrm{MeV}$ and width $\Gamma[T] =(140 \pm 13)~ \mathrm{MeV%
}$ of the tetraquark $T$, as well as its numerous decay channels explored in
this article are useful for ongoing and future experimental investigations
of fully heavy resonances.
\end{abstract}

\maketitle


\section{Introduction}

\label{sec:Intro}

Four-quark exotic mesons containing only heavy $\ b$ and/or $c$ quarks
recently became objects of experimental and theoretical studies.
Observations of four $X$ resonances by LHCb-ATLAS-CMS collaborations with
masses $6.2-7.3\ \mathrm{GeV}$ is one of important experimental achievements
of last years. These structures are supposedly four-quark mesons built of $c$
and $\overline{c}$ quarks \cite%
{LHCb:2020bwg,Bouhova-Thacker:2022vnt,CMS:2023owd}.

The discoveries of experimental groups inspired theoretical investigations
devoted to analysis of these resonances \cite%
{Zhang:2020xtb,Albuquerque:2020hio,Yang:2020wkh,Becchi:2020mjz,Becchi:2020uvq,Wang:2022xja, Faustov:2022mvs,Niu:2022vqp,Dong:2022sef,Yu:2022lak,Kuang:2023vac,Wang:2023kir,Dong:2020nwy,Liang:2021fzr}%
. They were usually considered as structures composed of four heavy quarks
with diquark-antidiquark or hadronic molecules organizations. These
resonances were studied in our works \cite%
{Agaev:2023wua,Agaev:2023ruu,Agaev:2023gaq,Agaev:2023rpj}, as well. In these
papers we calculated the masses and widths of various $cc\overline{c}%
\overline{c}$ states in the context of QCD sum rule method and compared the
obtained results with the measured parameters of the $X$ resonances. It is
worth to emphasize that, we used both the diquark-antidiquark and hadronic
molecule models. On basis of these calculations, we made suggestions about
possible structures of $X$ resonances. Thus, the resonance $X(6600)$ was
interpreted as a four-quark meson composed of axial-vector components \cite%
{Agaev:2023wua}. The light resonance $X(6200)$ may be considered as a
molecule $\eta _{c}\eta _{c}$ \cite{Agaev:2023ruu}, whereas $X(6900)$ is
presumably an admixture of diquark-antidiquark state and hadronic molecule $%
\chi _{c0}\chi _{c0}$ \cite{Agaev:2023ruu,Agaev:2023gaq}. The heaviest state
$X(7300)$ can be treated as mixing of the molecule $\chi _{c1}\chi _{c1}$
and first radial excitation of $X(6600)$ \cite{Agaev:2023rpj}.

The hidden charm-bottom exotic mesons, i.e., particles with contents $bc%
\overline{b}\overline{c}$ are another class of fully heavy four-quark
states. These particles were not yet observed in experiments, nevertheless
they attracted interests of researches. Their properties were studied in the
publications \cite%
{Faustov:2022mvs,Wu:2016vtq,Liu:2019zuc,Chen:2019vrj,Bedolla:2019zwg,Cordillo:2020sgc,Weng:2020jao,Yang:2021zrc,Hoffer:2024alv}%
, in which authors computed the masses of the tetraquarks $bc\overline{b}%
\overline{c}$ with different spin-parities by employing numerous methods.
Predictions obtained in these works in some aspects are controversial which
makes necessary further detailed analysis of these exotic mesons.

In our paper \cite{Agaev:2024wvp}, we explored the scalar particle $T_{%
\mathrm{bc\overline{b}\overline{c}}}=bc\overline{b}\overline{c}$ in the
context of the diquark-antidiquark model, and calculated the mass and width
of this state. The mass $m_{\mathrm{S}}=(12697\pm 90)~\mathrm{MeV}$ of $T_{%
\mathrm{bc\overline{b}\overline{c}}}$ demonstrates that it is unstable
against strong fall-apart processes, i.e., to decays when all constituent
quarks appear in two final mesons. These decays are the dominant channels
for transformation of $T_{\mathrm{bc\overline{b}\overline{c}}}$ to
conventional mesons. Besides, annihilations of $b$ and $\overline{b}$ quarks
in $T_{\mathrm{bc\overline{b}\overline{c}}}$ and creations of $DD$ meson
pairs are the alternative processes for such transformation \cite%
{Becchi:2020mjz, Becchi:2020uvq,Agaev:2023ara}. These decays can be
considered as subdominant channels for the tetraquark under discussion. The
width of the $T_{\mathrm{bc\overline{b}\overline{c}}}$ was evaluated by
taking into account its decay modes $\eta _{b}\eta _{c}$, $%
B_{c}^{+}B_{c}^{-} $, $B_{c}^{\ast +}B_{c}^{\ast -}$ and $%
B_{c}^{+}(1^{3}P_{0})B_{c}^{\ast -}$, as well as processes $T_{\mathrm{bc%
\overline{b}\overline{c}}}\rightarrow D^{+}D^{-}$, $D^{0}\overline{D}^{0}$, $%
D^{\ast }{}^{+}D^{\ast -}$, and $D^{\ast }{}^{0}\overline{D}^{\ast }{}^{0}$.

In the current work, we continue our investigations and explore features of
the axial-vector exotic meson $bc\overline{b}\overline{c}$. For simplicity,
throughout this paper, we label it as $T$. We plan to consider this particle
in a detailed form by computing its mass and full width. The mass $m$ and
current coupling $\Lambda $ of $T$ are calculated in the context of the
two-point sum rule (SR) approach \cite{Shifman:1978bx,Shifman:1978by}. It
turns out that $m$ exceeds the two-meson $\eta _{b}J/\psi $, $\eta
_{c}\Upsilon (1S)$, $B_{c}^{-}B_{c}^{\ast +}$, and $B_{c}^{+}B_{c}^{\ast -}$
thresholds. The subdominant processes generated by $b\overline{b}$
annihilation which will be studied are decays $T\rightarrow D^{0}\overline{D}%
^{\ast 0}$, $D^{\ast 0}\overline{D}^{0}$, $D^{\ast +}D^{-}$, $D^{+}D^{\ast
-} $. In the case $c\overline{c}\rightarrow q\overline{q}$ we will consider
decays $T\rightarrow B^{\ast +}B^{-}$, $B^{+}B^{\ast -}$, $B^{\ast 0}%
\overline{B}^{0}$, and $B^{0}\overline{B}^{\ast 0}$.

This article is structured in the following manner: In Sec.\ \ref{sec:Scalar}%
, we calculate the mass and current coupling of the axial-vector state $T$.
Result for $m$ allows us to fix kinematically possible decay channels of the
tetraquark $T$. In Sec.\ \ref{sec:ScalarWidths1}, we compute partial widths
of the dominant decay modes. To this end, we make use of the three-point sum
rule method which is necessary to evaluate the strong couplings at relevant
tetraquark-meson-meson vertices. The subdominant decay modes $T\rightarrow
D^{0}\overline{D}^{\ast 0}$, $D^{\ast 0}\overline{D}^{0}$, $D^{\ast +}D^{-}$%
, $D^{+}D^{\ast -}$ are analyzed in Sec.\ \ref{sec:ScalarWidths2}, where we
compute their widths. The decays $T\rightarrow B^{\ast +}B^{-}$, $%
B^{+}B^{\ast -}$, $B^{\ast 0}\overline{B}^{0}$, and $B^{0}\overline{B}^{\ast
0}$ are considered in Sec.\ \ref{sec:ScalarWidths3}. Here, we also determine
the full width of the axial-vector tetraquark $T$ . Section\ \ref{sec:Conc}
is reserved for our concluding notes.


\section{Spectroscopic parameters $m$ and $\Lambda $}

\label{sec:Scalar}
The mass $m$ and current coupling $\Lambda $ are important parameters of the
axial-vector tetraquark $T$, which can be calculated using the two-point sum
rule method \cite{Shifman:1978bx,Shifman:1978by}. We model this particle as
a diquark-antidiquark state made of a diquark $b^{T}C\gamma _{5}c$ and
antidiquark $\overline{b}\gamma _{\mu }C\overline{c}^{T}$. Then, the
relevant interpolating current has the form
\begin{eqnarray}
J_{\mu }(x) &=&b_{a}^{T}(x)C\gamma _{5}c_{b}(x)\left[ \overline{b}%
_{a}(x)\gamma _{\mu }C\overline{c}_{b}^{T}(x)\right.  \notag \\
&&\left. -\overline{b}_{b}(x)\gamma _{\mu }C\overline{c}_{a}^{T}(x)\right] ,
\label{eq:CR1}
\end{eqnarray}%
where $a$ and $b$ are the color indices and $C$ is the charge conjugation
matrix. The current $J_{\mu }(x)$ corresponds to the tetraquark with the
quantum numbers $J^{\mathrm{P}}=1^{+}$ and unfixed $C$ parity.

To find the sum rules for the mass $m$ and current coupling $\Lambda $, it
is necessary to investigate the correlation function%
\begin{equation}
\Pi _{\mu \nu }(p)=i\int d^{4}xe^{ipx}\langle 0|\mathcal{T}\{J_{\mu
}(x)J_{\nu }^{\dag }(0)\}|0\rangle ,  \label{eq:CF1}
\end{equation}%
with $\mathcal{T}$ \ being the time-ordered product of two currents. The SRs
for $m$ and $\Lambda $ are obtained by calculating the function $\Pi _{\mu
\nu }(p)$ both in terms of the physical parameters of the tetraquark $T$ and
heavy quark propagators. The function $\Pi _{\mu \nu }^{\mathrm{Phys}}(p)$
establishes the phenomenological side of the sum rules, whereas $\Pi _{\mu
\nu }^{\mathrm{OPE}}(p)$ forms their QCD components.

The correlation function $\Pi _{\mu \nu }^{\mathrm{Phys}}(p)$ is determined
by the formula
\begin{equation}
\Pi _{\mu \nu }^{\mathrm{Phys}}(p)=\frac{\langle 0|J_{\mu }|T(p,\epsilon
)\rangle \langle T(p,\epsilon )|J_{\nu }^{\dagger }|0\rangle }{m^{2}-p^{2}}%
+\cdots ,  \label{eq:Phys1}
\end{equation}%
where we have written down explicitly the contribution arising from the
ground-state particle, and denoted terms coming from higher resonances and
continuum states by the dots.

This expression can be recast into the more simple form
\begin{equation}
\Pi _{\mu \nu }^{\mathrm{Phys}}(p)=\frac{\Lambda ^{2}}{m^{2}-p^{2}}\left(
-g_{\mu \nu }+\frac{p_{\mu }p_{\nu }}{m^{2}}\right) +\cdots .
\label{eq:PhysSide1}
\end{equation}%
To derive Eq.\ (\ref{eq:PhysSide1}), we make use of the matrix element
\begin{equation}
\langle 0|J_{\mu }|T(p,\epsilon )\rangle =\Lambda \epsilon _{\mu }(p),
\label{eq:ME1}
\end{equation}%
where $\epsilon _{\mu }(p)$ is the polarization vector of the tetraquark $T$.

The correlation function $\Pi _{\mu \nu }^{\mathrm{Phys}}(p)$ is a sum of
two components with different Lorentz structures. For our purposes, it is
convenient to work with the term $\sim g_{\mu \nu }$, since it contains
contributions of only spin-$1$ particle. Then corresponding coefficient
function $\sim \Lambda ^{2}/(m^{2}-p^{2})$ is equal to the invariant
amplitude $\Pi ^{\mathrm{Phys}}(p^{2})$ which is required to find desired
SRs.

The correlator $\Pi _{\mu \nu }(p)$ should be calculated with fixed accuracy
using the quark-gluon degrees of freedom and operator product expansion ($%
\mathrm{OPE}$), which we denote as $\Pi _{\mu \nu }^{\mathrm{OPE}}(p)$. It
establishes the QCD side of the SRs and, as $\Pi _{\mu \nu }^{\mathrm{Phys}%
}(p)$, contains two Lorentz structures. The amplitude $\Pi ^{\mathrm{OPE}%
}(p^{2})$ corresponding to a term $g_{\mu \nu }$ in $\Pi _{\mu \nu }^{%
\mathrm{OPE}}(p)$ is the second ingredient in our analysis. This amplitude
contains a perturbative and a dimension four nonperturbative contribution:
The latter is proportional to $\langle \alpha _{s}G^{2}/\pi \rangle $.

The $\Pi _{\mu \nu }^{\mathrm{OPE}}(p)$ in terms of $b$ and $c$ quarks'
propagators is given by the expression
\begin{eqnarray}
&&\Pi _{\mu \nu }^{\mathrm{OPE}}(p)=i\int d^{4}xe^{ipx}\mathrm{Tr}\left[
\gamma _{5}\widetilde{S}_{b}^{aa^{\prime }}(x)\gamma _{5}S_{c}^{bb^{\prime
}}(x)\right]  \notag \\
&&\times \left\{ \mathrm{Tr}\left[ \gamma _{\mu }\widetilde{S}%
_{c}^{a^{\prime }b}(-x)\gamma _{\nu }S_{b}^{b^{\prime }a}(-x)\right] +%
\mathrm{Tr}\left[ \gamma _{\mu }\widetilde{S}_{c}^{b^{\prime }a}(-x)\right.
\right.  \notag \\
&&\left. \times \gamma _{\nu }S_{b}^{a^{\prime }}(-x)\right] -\mathrm{Tr}%
\left[ \gamma _{\mu }\widetilde{S}_{c}^{b^{\prime }b}(-x)\gamma _{\nu
}S_{b}^{a^{\prime }a}(-x)\right]  \notag \\
&&\left. -\mathrm{Tr}\left[ \gamma _{\mu }\widetilde{S}_{c}^{a^{\prime
}a}(-x)\gamma _{\nu }S_{b}^{b^{\prime }b}(-x)\right] \right\} ,
\label{eq:QCD1}
\end{eqnarray}%
where%
\begin{equation}
\widetilde{S}_{b(c)}(x)=CS_{b(c)}^{T}(x)C.  \label{eq:Prop}
\end{equation}%
In Eq.\ (\ref{eq:QCD1}) $S_{b(c)}(x)$ are propagators of the $b$ and $c$%
-quarks \cite{Agaev:2020zad}.

As usual, having equated the amplitudes $\Pi ^{\mathrm{OPE}}(p^{2})$ and $%
\Pi ^{\mathrm{Phys}}(p^{2})$, applied the Borel transformation and carried
out the continuum subtraction one derives the SRs for $m$ and $\Lambda $
\begin{equation}
m^{2}=\frac{\Pi ^{\prime }(M^{2},s_{0})}{\Pi (M^{2},s_{0})},  \label{eq:Mass}
\end{equation}%
and
\begin{equation}
\Lambda ^{2}=e^{m^{2}/M^{2}}\Pi (M^{2},s_{0}),  \label{eq:Coupl}
\end{equation}%
where $\Pi (M^{2},s_{0})$ is the amplitude $\Pi ^{\mathrm{OPE}}(p^{2})$
after the Borel transformation and continuum subtraction procedures. It
depends on the Borel and continuum subtraction parameters $M^{2}$ and $s_{0}$%
. In Eq.\ (\ref{eq:Mass}), we have introduced the short-hand notation $\Pi
^{\prime }(M^{2},s_{0})=d\Pi (M^{2},s_{0})/d(-1/M^{2})$.

The function $\Pi (M^{2},s_{0})$ can be expressed in terms of the two-point
spectral density $\rho ^{\mathrm{OPE}}(s)$ and nonperturbative function $\Pi
(M^{2})$
\begin{equation}
\Pi (M^{2},s_{0})=\int_{4\mathcal{M}^{2}}^{s_{0}}ds\rho ^{\mathrm{OPE}%
}(s)e^{-s/M^{2}}+\Pi (M^{2}),  \label{eq:InvAmp}
\end{equation}%
where $\mathcal{M}$ is equal to $m_{b}+m_{c}$. The spectral density $\rho ^{%
\mathrm{OPE}}(s)$ is equal to imaginary part of the amplitude $\Pi ^{\mathrm{%
OPE}}(p^{2})$ and consists of perturbative $\rho ^{\mathrm{pert.}}(s)$ and
nonperturbative $\rho ^{\mathrm{Dim4}}(s)$ terms. The function $\Pi (M^{2})$
is calculated directly from $\Pi _{\mu \nu }^{\mathrm{OPE}}(p)$ and contains
terms which are not enter to the spectral density. Explicit expressions of $%
\rho ^{\mathrm{OPE}}(s)$ and $\Pi (M^{2})$ are lengthy and not presented
here.

The sum rules given by Eqs.\ (\ref{eq:Mass}) and (\ref{eq:Coupl}) contain as
input parameters the masses of $b$ and $c$ quarks and gluon condensate:
Below, we write down their values used in our studies%
\begin{eqnarray}
&&m_{b}=4.18_{-0.02}^{+0.03}~\mathrm{GeV}\text{,\ }m_{c}=(1.27\pm 0.02)~%
\mathrm{GeV}\text{,}  \notag \\
&&\langle \alpha _{s}G^{2}/\pi \rangle =(0.012\pm 0.004)~\mathrm{GeV}^{4}.
\label{eq:GluonCond}
\end{eqnarray}

Other important quantities which should be fixed are the parameters $M^{2}$
and $s_{0}$. Their choice is not arbitrary but should meet standard
constraints of the sum rule method. One of the main requirements imposed on $%
M^{2}$ and $s_{0}$ is dominance of the pole contribution to extract the
physical parameters of the tetraquark $T$. A prevalence of the pole
contribution ($\mathrm{PC}$) implies fulfillment of the constraint $\mathrm{PC%
}\geq 0.5$, where
\begin{equation}
\mathrm{PC}=\frac{\Pi (M^{2},s_{0})}{\Pi (M^{2},\infty )}.  \label{eq:PC}
\end{equation}%
This restriction allows one to get the upper bound for $M^{2}$ as well. The
convergence of $\mathrm{OPE}$ is second important condition in the SR
analysis. Minimal value of $M^{2}$ is determined from this restriction. In
our case, there are two terms in the correlator $\Pi (M^{2},s_{0})$;
perturbative and dimension $4$ nonperturbative terms. Therefore, to ensure
the convergence of $\mathrm{OPE}$, we choose the minimal value of $M^{2}$ in
such a way that satisfies at least the constraint $|\Pi ^{\mathrm{Dim4}%
}(M^{2},s_{0})|=(0.05-0.15)\Pi (M^{2},s_{0})$. The stability of the
extracted $m$ and $\Lambda $ on the parameter $M^{2}$ is also among usual
conditions of SR studies.

\begin{figure}[h]
\includegraphics[width=8.5cm]{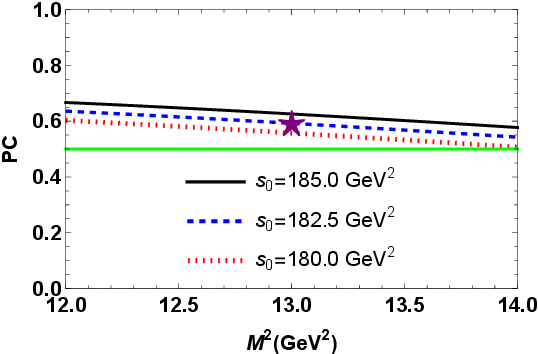}
\caption{$\mathrm{PC}$ as a function of $M^{2}$. The red star fixes the
point $M^{2}=13~\mathrm{GeV}^{2}$ and $s_{0}=182.5~\mathrm{GeV}^{2}$. }
\label{fig:PC}
\end{figure}

Our analysis proves that these SR restrictions are met by the working
intervals
\begin{equation}
M^{2}\in \lbrack 12,14]~\mathrm{GeV}^{2},\ s_{0}\in \lbrack 180,185]~\mathrm{%
GeV}^{2}.  \label{eq:Wind1}
\end{equation}%
Thus, on the average in $s_{0}$, at maximal and minimal values of $M^{2}$
the pole contribution is $\mathrm{PC}\approx 0.54$, and $\approx 0.64$,
respectively. The nonperturbative term is positive and at $M^{2}=12~\mathrm{%
GeV}^{2}$ forms $<1\%$ of the whole result. In Fig.\ \ref{fig:PC},
dependence of $\ \mathrm{PC}$ on the Borel parameter is plotted. It is
evident that it overshoots $0.5$ for all $M^{2}$ and $s_{0}$.

\begin{widetext}

\begin{figure}[h!]
\begin{center}
\includegraphics[totalheight=6cm,width=8cm]{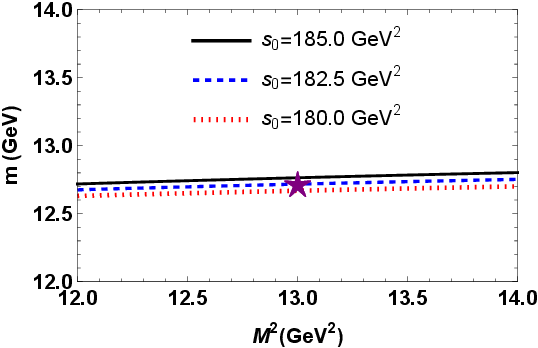}
\includegraphics[totalheight=6cm,width=8cm]{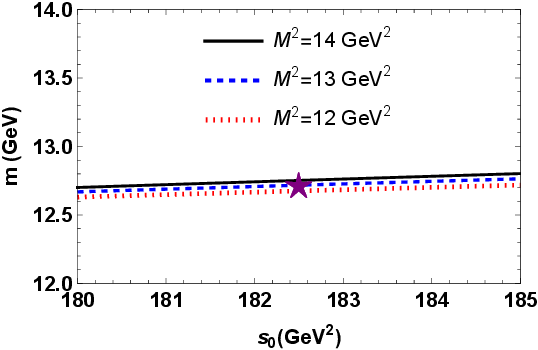}
\end{center}
\caption{Dependence of the mass $m$ on the Borel  $M^{2}$ (left panel), and continuum threshold $s_0$ (right panel) parameters.}
\label{fig:Mass}
\end{figure}

\end{widetext}

We calculate the parameters $m$ and $\Lambda $ as their mean values in
regions Eq.\ (\ref{eq:Wind1}) and get
\begin{eqnarray}
m &=&(12715\pm 90)~\mathrm{MeV},  \notag \\
\Lambda &=&(2.31\pm 0.26)~\mathrm{GeV}^{5}.  \label{eq:Result1}
\end{eqnarray}%
These results correspond to SR predictions at the point $M^{2}=13~\mathrm{GeV%
}^{2}$ and $s_{0}=182.5~\mathrm{GeV}^{2}$, where $\mathrm{PC}\approx 0.59$,
which guarantees the dominance of $\mathrm{PC}$ in the extracted parameters.
In Fig.\ \ref{fig:Mass}, we demonstrate $m$ as a function of $M^{2}$ and $%
s_{0}$.

It is interesting to compare our result for the mass of the axial-vector
tetraquark $T$ with predictions made in other publications. As it has been
noted above, the masses of exotic mesons $bc\overline{b}\overline{c}$ with
different spin-parities $J^{\mathrm{PC}}$ were calculated in numerous works
\cite%
{Faustov:2022mvs,Wu:2016vtq,Liu:2019zuc,Chen:2019vrj,Bedolla:2019zwg,Cordillo:2020sgc,Weng:2020jao,Yang:2021zrc,Hoffer:2024alv}%
. In these articles authors used various methods and sometimes their results
differ from each other significantly. In fact, in the framework of SR method
the tetraquarks $bc\overline{b}\overline{c}$ were studied in Ref.\ \cite%
{Yang:2021zrc}, in which the mass of the axial-vector state was estimated
within limits $12300-12360~\mathrm{MeV}$. The relativistic quark model led
to predictions $12826-12831~\mathrm{MeV}$ \cite{Faustov:2022mvs}. In the
color-magnetic interaction model the mass of the tetraquark $T$ was found
equal to $13478-13599~\mathrm{MeV}$ \cite{Wu:2016vtq}, whereas the potential
model gave $12881-13020~\mathrm{MeV}$ \cite{Liu:2019zuc}. In the extended
chromomagnetic model, the authors obtained $12671.7~\mathrm{MeV}$ \cite%
{Weng:2020jao}. Recently, the masses of hidden charm-bottom tetraquarks were
computed in Ref.\ \cite{Hoffer:2024alv}, where the authors fixed $%
11130-11460~\mathrm{MeV}$ as boundaries for the mass of the axial-vector
diquark-antidiquark state $T$.

Our prediction is close to the results of Refs.\ \cite%
{Faustov:2022mvs,Weng:2020jao} and exceeds ones from Refs.\ \cite%
{Yang:2021zrc,Hoffer:2024alv}. The mass of this particle makes possible its
dissociations to two-meson states. For instance, decays to $\eta _{b}J/\psi $
and $\eta _{c}\Upsilon (1S)$ mesons are kinematically allowed processes,
because corresponding thresholds amount to $12496~\mathrm{MeV}$ and $12445~%
\mathrm{MeV}$, respectively. These and other processes are explored in the
next sections.

\section{Decays of $T$: Dissociation processes}

\label{sec:ScalarWidths1}

In this section we consider decays of $T$ to meson pairs, in which all
constituent quarks participate in produced final-state particles. It is
evident that a four-quark meson built of $bc\overline{b}\overline{c}$ quarks
may dissociate to ordinary $(\overline{c}c$, $\overline{b}b)$ and $(%
\overline{c}b$, $\overline{b}c)$ mesons with appropriate charges and
spin-parities, if the mass of this particle exceeds masses of generated
mesons. In the case under consideration, the axial-vector tetraquark $T$
with the mass $12715~\mathrm{MeV}$ can easily dissociate to the pairs $\eta
_{b}J/\psi $, $\eta _{c}\Upsilon (1S)$, $B_{c}^{-}B_{c}^{\ast +}$, and $%
B_{c}^{+}B_{c}^{\ast -}$. This section is devoted to investigation of these
processes.


\subsection{Decay $T\rightarrow \protect\eta _{b}J/\protect\psi $}


We start from analysis of the decay $T\rightarrow \eta _{b}J/\psi $. To
calculate the partial width of this process, one needs to determine the
strong coupling $g_{1}$ of the involved particles at the vertex $T\eta
_{b}J/\psi $. To find it, we consider the three-point correlation function
\begin{eqnarray}
\Pi _{\mu \nu }^{1}(p,p^{\prime }) &=&i^{2}\int d^{4}xd^{4}ye^{ip^{\prime
}y}e^{-ipx}\langle 0|\mathcal{T}\{J^{\eta _{b}}(y)  \notag \\
&&\times J_{\nu }^{J/\psi }(0)J_{\mu }^{\dagger }(x)\}|0\rangle ,
\label{eq:CF3}
\end{eqnarray}%
where
\begin{equation}
J^{\eta _{b}}(x)=\overline{b}_{i}(x)i\gamma _{5}b_{i}(x),\ J_{\nu }^{J/\psi
}(x)=\overline{c}_{j}(x)\gamma _{\nu }c_{j}(x),  \label{eq:CR3}
\end{equation}%
are the interpolating currents of the quarkonia $\eta _{b}$ and $J/\psi $,
respectively.

Analysis of correlator $\Pi _{\mu \nu }^{1}(p,p^{\prime })$ allows us to
derive the SR for the form factor $g_{1}(q^{2})$, which at the mass shell $%
q^{2}=m_{J/\psi }^{2}$ is equal to the strong coupling $g_{1}$. The sum rule
for $g_{1}(q^{2})$ can be obtained using well-known prescriptions of the
method. To this end, we write $\Pi _{\mu \nu }^{1}(p,p^{\prime })$ using
matrix elements of the tetraquark $T$, and mesons $\eta _{b}$ and $J/\psi $
\begin{eqnarray}
&&\Pi _{\mu \nu }^{1\mathrm{Phys}}(p,p^{\prime })=\frac{\langle 0|J^{\eta
_{b}}|\eta _{b}(p^{\prime })\rangle }{p^{\prime 2}-m_{\eta _{b}}^{2}}\frac{%
\langle 0|J_{\nu }^{J/\psi }|J/\psi (q,\varepsilon )\rangle }{%
q^{2}-m_{J/\psi }^{2}}  \notag \\
&&\times \langle \eta _{b}(p^{\prime })J/\psi (q,\varepsilon )|T(p,\epsilon
)\rangle \frac{\langle T(p,\epsilon )|J_{\mu }^{\dagger }|0\rangle }{%
p^{2}-m^{2}}+\cdots .  \notag \\
&&  \label{eq:CF5}
\end{eqnarray}%
Here $m_{\eta _{b}}=(9398.7\pm 2.0)~\mathrm{MeV}$ and $m_{J/\psi
}=(3096.900\pm 0.006)~\mathrm{MeV}$ are the masses of $\eta _{b}$ and $%
J/\psi $ mesons \cite{PDG:2022}, respectively. The term in Eq.\ (\ref{eq:CF5}%
) is the contribution of the ground-state particles: Contributions to $\Pi
_{\mu \nu }^{1\mathrm{Phys}}(p,p^{\prime })$ arising from higher resonances
and continuum states are shown by the ellipses.

The function $\Pi _{\mu \nu }^{1\mathrm{Phys}}(p,p^{\prime })$ can be
computed by employing the matrix elements
\begin{eqnarray}
&&\langle 0|J^{\eta _{b}}|\eta _{b}(p^{\prime })\rangle =\frac{f_{\eta
_{b}}m_{\eta _{b}}^{2}}{2m_{b}},  \notag \\
&&\langle 0|J_{\nu }^{J/\psi }|J/\psi (q,\varepsilon )\rangle =f_{J/\psi
}m_{J/\psi }\varepsilon _{\nu }(q),  \label{eq:ME2}
\end{eqnarray}%
where $f_{\eta _{b}}=724~\mathrm{MeV}$ and $f_{J/\psi }=409~\mathrm{MeV}$
are the decay constants of the mesons $\eta _{b}$ and $J/\psi $,
respectively. In Eq.\ (\ref{eq:ME2}) $\varepsilon _{\nu }(q)$ is the
polarization vector of the meson $J/\psi $. The vertex $T\eta _{b}J/\psi $
is given by the following formula
\begin{eqnarray}
&&\langle \eta _{b}(p^{\prime })J/\psi (q,\varepsilon )|T(p,\epsilon
)\rangle =g_{1}(q^{2})\left[ p\cdot p^{\prime }\epsilon \cdot \varepsilon
^{\ast }\right.  \notag \\
&&\left. -p\cdot \varepsilon ^{\ast }p^{\prime }\cdot \epsilon \right] .
\label{eq:ME3}
\end{eqnarray}%
As a result, the correlator $\Pi _{\mu \nu }^{1\mathrm{Phys}}(p,p^{\prime })$
becomes equal to
\begin{eqnarray}
&&\Pi _{\mu \nu }^{1\mathrm{Phys}}(p,p^{\prime })=g_{1}(q^{2})\frac{\Lambda
f_{\eta _{b}}m_{\eta _{b}}^{2}f_{J/\psi }m_{J/\psi }}{2m_{b}\left(
p^{2}-m^{2}\right) }  \notag \\
&&\times \frac{1}{\left( p^{\prime 2}-m_{\eta _{b}}^{2}\right)
(q^{2}-m_{J/\psi }^{2})}\left[ \frac{m^{2}+m_{\eta _{b}}^{2}-q^{2}}{2}g_{\mu
\nu }\right.  \notag \\
&&-\frac{m^{2}+m_{\eta _{b}}^{2}-q^{2}}{2m_{J/\psi }^{2}}(p_{\mu }p_{\nu
}-p_{\mu }p_{\nu }^{\prime })-\frac{m^{2}}{m_{J/\psi }^{2}}p_{\mu }^{\prime
}p_{\nu }^{\prime }  \notag \\
&&\left. +\frac{m^{2}-m_{J/\psi }^{2}}{m_{J/\psi }^{2}}p_{\mu }^{\prime
}p_{\nu }\right] +\cdots .  \label{eq:CF6}
\end{eqnarray}%
The correlation function $\Pi _{\mu \nu }^{1\mathrm{Phys}}(p,p^{\prime })$
contains different Lorentz structures. To continue, we choose the term which
is proportional to $g_{\mu \nu }$, and denotes the corresponding invariant
amplitude by $\Pi _{1}^{\mathrm{Phys}}(p^{2},p^{\prime 2},q^{2})$.

The same function, $\Pi _{\mu \nu }^{1}(p,p^{\prime })$, expressed by means
of the heavy quark propagators reads
\begin{eqnarray}
&&\Pi _{\mu \nu }^{1\mathrm{OPE}}(p,p^{\prime })=i\int
d^{4}xd^{4}ye^{ip^{\prime }y}e^{-ipx}\left\{ \mathrm{Tr}\left[ \gamma
_{5}S_{b}^{ja}(y-x)\right. \right.  \notag \\
&&\left. \times \gamma _{5}\widetilde{S}_{c}^{ib}(-x)\gamma _{\nu }%
\widetilde{S}_{c}^{bi}(x)\gamma _{\mu }S_{b}^{aj}(x-y)\right]  \notag \\
&&\left. -\mathrm{Tr}\left[ \gamma _{5}S_{b}^{ja}(y-x)\gamma _{5}\widetilde{S%
}_{c}^{ib}(-x)\gamma _{\nu }\widetilde{S}_{c}^{ai}(x)\gamma _{\mu
}S_{b}^{bj}(x-y)\right] \right\} .  \notag \\
&&  \label{eq:QCDside2}
\end{eqnarray}%
We label by $\Pi ^{1\mathrm{OPE}}(p^{2},p^{\prime 2},q^{2})$ the invariant
amplitude which corresponds to the term $\sim g_{\mu \nu }$ in $\Pi _{\mu
\nu }^{1\mathrm{OPE}}(p,p^{\prime })$, and utilize it to determine the SR
for the form factor $g_{1}(q^{2})$%
\begin{eqnarray}
&&g_{1}(q^{2})=\frac{4m_{b}}{\Lambda f_{\eta _{b}}m_{\eta _{b}}^{2}f_{J/\psi
}m_{J/\psi }}\frac{q^{2}-m_{J/\psi }^{2}}{m^{2}+m_{\eta _{b}}^{2}-q^{2}}
\notag \\
&&\times e^{m^{2}/M_{1}^{2}}e^{m_{\eta _{b}}^{2}/M_{2}^{2}}\Pi _{1}(\mathbf{M%
}^{2},\mathbf{s}_{0},q^{2}),  \label{eq:SRCoup2}
\end{eqnarray}%
where
\begin{eqnarray}
&&\Pi _{1}(\mathbf{M}^{2},\mathbf{s}_{0},q^{2})=\int_{4\mathcal{M}%
^{2}}^{s_{0}}ds\int_{4m_{b}^{2}}^{s_{0}^{\prime }}ds^{\prime }\rho
_{1}(s,s^{\prime },q^{2})  \notag \\
&&\times e^{-s/M_{1}^{2}}e^{-s^{\prime }/M_{2}^{2}},  \label{eq:AS1}
\end{eqnarray}%
is the function $\Pi ^{1\mathrm{OPE}}(p^{2},p^{\prime 2},q^{2})$ after Borel
transformations over the variables $-p^{2}$, $-p^{\prime 2}$ and relevant
continuum subtractions. It is written down in terms of the spectral density $%
\rho _{1}(s,s^{\prime },q^{2})$.

The Borel $\mathbf{M}^{2}=(M_{1}^{2},M_{2}^{2})$ and continuum threshold
parameters $\mathbf{s}_{0}=(s_{0},s_{0}^{\prime })$ in Eq.\ (\ref{eq:AS1})
correspond to different particles, and should satisfy usual constraints of
the sum rule method. Our analysis confirms that $M_{1}^{2}$ and $s_{0}$
chosen from Eq.\ (\ref{eq:Wind1}), and parameters $(M_{2}^{2},\
s_{0}^{\prime })$ of the $\eta _{b}$ channel varying within limits
\begin{equation}
M_{2}^{2}\in \lbrack 10,12]~\mathrm{GeV}^{2},\ s_{0}^{\prime }\in \lbrack
95,99]~\mathrm{GeV}^{2},  \label{eq:Wind3}
\end{equation}%
meet all required condisions. It should be noted that$\ \sqrt{s_{0}^{\prime }%
}$ is restricted by the mass $9.999\ \mathrm{GeV}$ of the excited $\eta
_{b}(2S)$ meson.

Reliable predictions for $g_{1}(q^{2})$ in the context of the SR method can
be extracted in the Euclidean region $q^{2}<0$. But the coupling $g_{1}$ has
to be fixed at the mass shell $q^{2}=m_{J/\psi }^{2}$. To evade this
obstacle, we introduce a variable $Q^{2}=-q^{2}$ and employ $g_{1}(Q^{2})$
for the obtained new function. Afterwards, we introduce an extrapolating
function $\mathcal{G}_{1}(Q^{2})$ which for $Q^{2}>0$ leads to the SR data,
but can be extended to the region of negative $Q^{2}$ . For these purposes,
we employ the function
\begin{equation}
\mathcal{G}_{i}(Q^{2})=\mathcal{G}_{i}^{0}\mathrm{\exp }\left[ a_{i}^{1}%
\frac{Q^{2}}{m^{2}}+a_{i}^{2}\left( \frac{Q^{2}}{m^{2}}\right) ^{2}\right] ,
\end{equation}%
where $\mathcal{G}_{i}^{0}$, $a_{i}^{1}$, and $a_{i}^{2}$ are fitting
constants.

\begin{figure}[h]
\includegraphics[width=8.5cm]{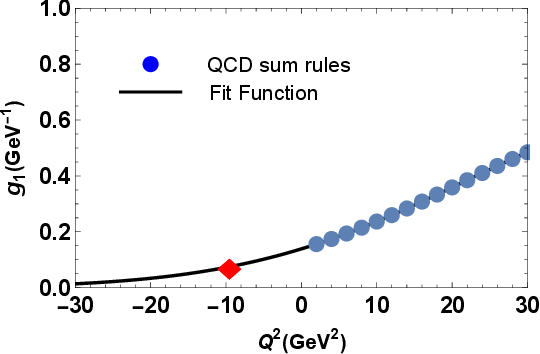}
\caption{QCD data and fit function for the form factor $g_{1}(Q^{2})$. The
diamond fix the point $Q^{2}=-m_{J/\protect\psi}^{2}$ where $g_{1}$ has been
estimated. }
\label{fig:Fit}
\end{figure}

We perform the sum rule computations in the region $Q^{2}=2-30~\mathrm{GeV}%
^{2}$, results of which are depicted in Fig.\ \ref{fig:Fit}. Here, we show
also the function $\mathcal{G}_{1}(Q^{2})$ with parameters $\mathcal{G}%
_{1}^{0}=0.14~\mathrm{GeV}^{-1}$, $a_{1}^{1}=9.55$, and $a_{1}^{2}=-15.19$:
as is seen, there is nice agreement between the SR data and $\mathcal{G}%
_{1}(Q^{2})$.

For $g_{1}$ we find
\begin{equation}
g_{1}\equiv \mathcal{G}_{1}(-m_{J/\psi }^{2})=(7.48\pm 0.89)\times 10^{-2}\
\mathrm{GeV}^{-1}.
\end{equation}%
The width of the decay $T\rightarrow \eta _{b}J/\psi $ can be found by means
of the formula%
\begin{equation}
\Gamma \left[ T\rightarrow \eta _{b}J/\psi \right] =g_{1}^{2}\frac{\lambda
_{1}}{24\pi }|M_{1}|^{2},  \label{eq:PDw2}
\end{equation}%
where
\begin{eqnarray}
|M_{1}|^{2} &=&\frac{1}{4m_{J/\psi }^{2}}\left[ m^{6}-2m^{4}m_{\eta
_{b}}^{2}+2(m_{J/\psi }^{3}-m_{\eta _{b}}^{2}m_{J/\psi })^{2}\right.  \notag
\\
&&\left. +m^{2}(m_{\eta _{b}}^{4}+6m_{\eta _{b}}^{2}m_{J/\psi
}^{2}-3m_{J/\psi }^{4})\right] .
\end{eqnarray}%
It is worth noting that in Eq.\ (\ref{eq:PDw2}) $\lambda _{1}=\lambda
(m,m_{\eta _{b}},m_{\eta _{c}})$ is equal to
\begin{equation}
\lambda (x,y,z)=\frac{\sqrt{%
x^{4}+y^{4}+z^{4}-2(x^{2}y^{2}+x^{2}z^{2}+y^{2}z^{2})}}{2x}.
\end{equation}%
Then, we get
\begin{equation}
\Gamma \left[ T\rightarrow \eta _{b}J/\psi \right] =(21.5\pm 5.5)~\mathrm{MeV%
}.  \label{eq:DW1}
\end{equation}

\subsection{Process $T\rightarrow \protect\eta _{c}\Upsilon (1S)$}


The partial width of the process $T\rightarrow \eta _{c}\Upsilon $ is
determined by the strong coupling $g_{2}$ at the vertex $T\eta _{c}\Upsilon $%
. In the context of the SR method the form factor $g_{2}(q^{2})$ can be
computed by analyzing the following three-point correlation function%
\begin{eqnarray}
\Pi _{\mu \nu }^{2}(p,p^{\prime }) &=&i^{2}\int d^{4}xd^{4}ye^{ip^{\prime
}y}e^{-ipx}\langle 0|\mathcal{T}\{J_{\nu }^{\Upsilon }(y)  \notag \\
&&\times J^{\eta _{c}}(0)J_{\mu }^{\dagger }(x)\}|0\rangle .  \label{eq:CF7}
\end{eqnarray}%
The interpolating currents of the mesons $\eta _{c}$ and $\Upsilon $ in Eq.\
(\ref{eq:CF7}) are
\begin{equation}
J^{\eta _{c}}(x)=\overline{c}_{i}(x)i\gamma _{5}c_{i}(x),\ J_{\nu
}^{\Upsilon }(x)=\overline{b}_{j}(x)\gamma _{\nu }b_{j}(x).  \label{eq:C3}
\end{equation}

The matrix elements necessary to calculate $\Pi _{\mu \nu }^{2\mathrm{Phys}%
}(p,p^{\prime })$ are
\begin{eqnarray}
&&\langle 0|J_{\nu }^{\Upsilon }|\Upsilon (p^{\prime },\varepsilon )\rangle
=f_{\Upsilon }m_{\Upsilon }\varepsilon _{\nu }(q),  \notag \\
&&\langle 0|J^{\eta _{c}}|\eta _{c}(q)\rangle =\frac{f_{\eta _{c}}m_{\eta
_{c}}^{2}}{2m_{c}},  \label{eq:ME4}
\end{eqnarray}%
where $f_{\eta _{c}}$, $m_{\eta _{c}}$ and $f_{\Upsilon }$, $m_{\Upsilon }$
are decay constants and masses of the mesons $\eta _{c}$ and $\Upsilon $.
The vertex $T\eta _{c}\Upsilon $ is given by the expression
\begin{eqnarray}
&&\langle \eta _{c}(q)\Upsilon (p^{\prime },\varepsilon )|T(p,\epsilon
)\rangle =g_{2}(q^{2})\left[ p\cdot q\epsilon \cdot \varepsilon ^{\ast
}\right.  \notag \\
&&\left. -p\cdot \varepsilon ^{\ast }q\cdot \epsilon \right] .
\end{eqnarray}%
Then the correlator $\Pi _{\mu \nu }^{2\mathrm{Phys}}(p,p^{\prime })$
becomes equal to
\begin{eqnarray}
&&\Pi _{\mu \nu }^{2\mathrm{Phys}}(p,p^{\prime })=g_{2}(q^{2})\frac{\Lambda
f_{\eta _{c}}m_{\eta _{c}}^{2}f_{\Upsilon }m_{\Upsilon }}{2m_{c}\left(
p^{2}-m^{2}\right) \left( p^{\prime 2}-m_{\Upsilon }^{2}\right) }  \notag \\
&&\times \frac{1}{(q^{2}-m_{\eta _{c}}^{2})}\left[ \frac{m^{2}-m_{\Upsilon
}^{2}+q^{2}}{2}g_{\mu \nu }-p_{\mu }p_{\nu }\right.  \notag \\
&&\left. +p_{\mu }^{\prime }p_{\nu }-\frac{m^{2}}{m_{\Upsilon }^{2}}p_{\mu
}^{\prime }p_{\nu }^{\prime }+\frac{m^{2}+m_{\Upsilon }^{2}-q^{2}}{%
2m_{\Upsilon }^{2}}p_{\mu }p_{\nu }^{\prime }\right]  \notag \\
&&+\cdots .
\end{eqnarray}

The QCD side of the sum rule $\Pi _{\mu \nu }^{2\mathrm{OPE}}(p,p^{\prime })$
is determined by the formula
\begin{eqnarray}
&&\Pi _{\mu \nu }^{2\mathrm{OPE}}(p,p^{\prime })=i\int
d^{4}xd^{4}ye^{ip^{\prime }y}e^{-ipx}\left\{ \mathrm{Tr}\left[ \gamma
_{5}S_{c}^{ib}(-x)\right. \right.  \notag \\
&&\left. \times \gamma _{5}\widetilde{S}_{b}^{ja}(y-x)\gamma _{\nu }%
\widetilde{S}_{b}^{aj}(x-y)\gamma _{\mu }S_{c}^{bi}(x)\right]  \notag \\
&&\left. -\mathrm{Tr}\left[ \gamma _{5}S_{c}^{ib}(-x)\gamma _{5}\widetilde{S}%
_{b}^{ja}(y-x)\gamma _{\nu }\widetilde{S}_{b}^{bj}(x-y)\gamma _{\mu
}S_{c}^{ai}(x)\right] \right\} .  \notag \\
&&
\end{eqnarray}%
The functions $\Pi _{\mu \nu }^{2\mathrm{Phys}}(p,p^{\prime })$ and $\Pi
_{\mu \nu }^{2\mathrm{OPE}}(p,p^{\prime })$ have the same Lorentz
structures. For our studies, we consider terms $\sim g_{\mu \nu }$ and use
corresponding invariant amplitudes to find the SR for $g_{2}(q^{2})$

\begin{eqnarray}
&&g_{2}(q^{2})=\frac{4m_{c}}{\Lambda f_{\eta _{c}}m_{\eta
_{c}}^{2}f_{\Upsilon }m_{\Upsilon }}\frac{q^{2}-m_{\eta _{c}}^{2}}{%
m^{2}-m_{\Upsilon }^{2}+q^{2}}  \notag \\
&&\times e^{m^{2}/M_{1}^{2}}e^{m_{\Upsilon }^{2}/M_{2}^{2}}\Pi _{2}(\mathbf{M%
}^{2},\mathbf{s}_{0},q^{2}),  \label{eq:SRCoup}
\end{eqnarray}%
where $\Pi _{2}(\mathbf{M}^{2},\mathbf{s}_{0},q^{2})$ is Borel transformed
and subtracted amplitude $\Pi _{2}^{\mathrm{OPE}}(p^{2},p^{\prime 2},q^{2})$%
:
\begin{eqnarray}
&&\Pi _{2}(\mathbf{M}^{2},\mathbf{s}_{0},q^{2})=\int_{4\mathcal{M}%
^{2}}^{s_{0}}ds\int_{4m_{b}^{2}}^{s_{0}^{\prime }}ds^{\prime }\rho
_{2}(s,s^{\prime },q^{2})  \notag \\
&&\times e^{-s/M_{1}^{2}}e^{-s^{\prime }/M_{2}^{2}}.
\end{eqnarray}

Remaining manipulations are standard ones and have been explained above. In
numerical analysis, for the masses of the mesons $\eta _{c}$ and $\Upsilon
(1S)$, we use $m_{\eta _{c}}=(2984.1\pm 0.4)~\mathrm{MeV}$, $m_{\Upsilon
}=9460.40(09)(04)~\mathrm{MeV}\ $ from PDG \cite{PDG:2022}. The decay
constant $f_{\eta _{c}}=(421\pm 35)~\mathrm{MeV}$ was extracted from SR
analysis \cite{Veliev:2010vd}, whereas for $f_{\Upsilon }$ we use its
experimental value $(708\pm 8)~\mathrm{MeV}$ borrowed from Ref.\ \cite%
{Lakhina:2006vg}. We have utilized also the following windows for $M_{2}^{2}$%
, and $s_{0}^{\prime }$ in the $\Upsilon $ channel
\begin{equation}
M_{2}^{2}\in \lbrack 10,12]~\mathrm{GeV}^{2},\ s_{0}^{\prime }\in \lbrack
98,100]~\mathrm{GeV}^{2}.
\end{equation}

The extrapolating function $\mathcal{G}_{2}(Q^{2})$ has the parameters: $%
\mathcal{G}_{2}^{0}=0.42~\mathrm{GeV}^{-1}$, $a_{2}^{1}=10.64$, and $%
a_{2}^{2}=-8.09$. As a result, the strong coupling $g_{2}$ amounts to
\begin{equation}
g_{2}\equiv \mathcal{G}_{2}(-m_{\eta _{c}}^{2})=(2.3\pm 0.3)\times 10^{-1}\
\mathrm{GeV}^{-1}.
\end{equation}%
The width of the decay $T\rightarrow \eta _{c}\Upsilon (1S)$ is
\begin{equation}
\Gamma \left[ T\rightarrow \eta _{c}\Upsilon (1S)\right] =(24.6\pm 6.3)~%
\mathrm{MeV}.  \label{eq:DW2}
\end{equation}


\subsection{Modes $T\rightarrow B_{c}^{\ast +}B_{c}^{-}$ and $B_{c}^{\ast
-}B_{c}^{+}$}


The processes $T\rightarrow B_{c}^{\ast +}B_{c}^{-}$ and $B_{c}^{\ast
-}B_{c}^{+}$ are also allowed decay modes of the axial-vector
diquark-antidiquark state $T$. Analysis of these processes do not differ
considerably from ones studied in the previous subsections. Therefore, below we
present principal formulas and final results.

The correlation function which should be considered in the case of the decay
$T\rightarrow B_{c}^{\ast +}B_{c}^{-}$ is%
\begin{eqnarray}
\Pi _{\mu \nu }^{3}(p,p^{\prime }) &=&i^{2}\int d^{4}xd^{4}ye^{ip^{\prime
}y}e^{-ipx}\langle 0|\mathcal{T}\{J_{\nu }^{B_{c}^{\ast }}(y)  \notag \\
&&\times J^{B_{c}}(0)J_{\mu }^{\dagger }(x)\}|0\rangle .
\end{eqnarray}%
Here, $J_{\nu }^{B_{c}^{\ast }}(x)$ and $J^{B_{c}}(x)$ are the interpolating
currents of $B_{c}^{\ast +}$ and $B_{c}^{-}$ mesons, which are determined by
the expressions%
\begin{equation}
J_{\nu }^{B_{c}^{\ast }}(x)=\overline{b}_{i}(x)\gamma _{\nu }c_{i}(x),\
J^{B_{c}}(x)=\overline{c}_{j}(x)i\gamma _{5}b_{j}(x).
\end{equation}

In terms of the matrix elements and physical parameters of the involved
particles the correlator $\Pi _{\mu \nu }^{3}(p,p^{\prime })$ acquires the
following form
\begin{eqnarray}
&&\Pi _{\mu \nu }^{3\mathrm{Phys}}(p,p^{\prime })=\frac{\langle 0|J_{\nu
}^{B_{c}^{\ast }}|B_{c}^{\ast +}(p^{\prime },\varepsilon )\rangle }{%
p^{\prime 2}-m_{B_{c}^{\ast }}^{2}}\frac{\langle
0|J^{B_{c}}|B_{c}^{-}(q)\rangle }{q^{2}-m_{B_{c}}^{2}}  \notag \\
&&\times \langle B_{c}^{\ast +}(p^{\prime },\varepsilon
)B_{c}^{-}(q)|T(p,\epsilon )\rangle \frac{\langle T(p,\epsilon )|J^{\dagger
}|0\rangle }{p^{2}-m^{2}}+\cdots ,  \notag \\
&&
\end{eqnarray}%
where $m_{B_{c}^{\ast }}$ and $m_{B_{c}}$ are the masses of the mesons $%
B_{c}^{\ast +}$ and $B_{c}^{-}$, respectively. As the mass of the $B_{c}$
meson, we employ its experimental value $m_{B_{c}}=6274.47(27)(17)~\mathrm{%
MeV}$. The mass $m_{B_{c}^{\ast }}=6338~\mathrm{MeV}$ of the meson $%
B_{c}^{\ast }$ is theoretical prediction from Ref.\ \cite{Godfrey:2004ya}.
Subsequent calculations are carried out using the matrix elements
\begin{eqnarray}
&&\langle 0|J_{\nu }^{B_{c}^{\ast }}|B_{c}^{\ast +}(p^{\prime },\varepsilon
)\rangle =f_{B_{c}^{\ast }}m_{B_{c}^{\ast }}\varepsilon _{\nu }(p^{\prime }),
\notag \\
&&\langle 0|J^{B_{c}}|B_{c}^{-}(q)\rangle =\frac{f_{B_{c}}m_{B_{c}}^{2}}{%
m_{b}+m_{c}}.
\end{eqnarray}%
Here, $f_{B_{c}^{\ast }}$ and $\varepsilon _{\nu }(p^{\prime })$ are the
decay constant and polarization vector of $B_{c}^{\ast +}$, whereas $%
f_{B_{c}}$ denotes the decay constant of the meson $B_{c}^{-}$. The vertex $%
TB_{c}^{\ast +}B_{c}^{-}$ is considered in the form%
\begin{eqnarray}
&&\langle B_{c}^{\ast +}(p^{\prime },\varepsilon )B_{c}^{-}(q)|T(p,\epsilon
)\rangle =g_{3}(q^{2})\left[ p\cdot q\epsilon \cdot \varepsilon ^{\ast
}\right.  \notag \\
&&\left. -p\cdot \varepsilon ^{\ast }q\cdot \epsilon \right] .
\end{eqnarray}%
Then $\Pi _{\mu \nu }^{3}(p,p^{\prime })$ calculated by utilizing the
physical parameters of the involved particles is given by the expression
\begin{eqnarray}
&&\Pi _{\mu \nu }^{3\mathrm{Phys}}(p,p^{\prime })=\frac{g_{3}(q^{2})\Lambda
f_{B_{c}^{\ast }}m_{B_{c}^{\ast }}f_{B_{c}}m_{B_{c}}^{2}}{%
(m_{b}+m_{c})\left( p^{2}-m^{2}\right) \left( p^{\prime 2}-m_{B_{c}^{\ast
}}^{2}\right) }  \notag \\
&&\times \frac{1}{(q^{2}-m_{B_{c}}^{2})}\left[ \frac{m^{2}-m_{B_{c}^{\ast
}}^{2}+q^{2}}{2}g_{\mu \nu }-p_{\mu }p_{\nu }+p_{\mu }^{\prime }p_{\nu
}\right.  \notag \\
&&\left. -\frac{m^{2}}{m_{B_{c}^{\ast }}^{2}}p_{\mu }^{\prime }p_{\nu
}^{\prime }+\frac{m^{2}+m_{B_{c}^{\ast }}^{2}-q^{2}}{2m_{B_{c}^{\ast }}^{2}}%
p_{\mu }p_{\nu }^{\prime }\right] +\cdots .
\end{eqnarray}%
The function $\Pi _{\mu \nu }^{3}(p,p^{\prime })$ computed in terms of the
quark propagators reads%
\begin{eqnarray}
&&\Pi _{\mu \nu }^{3\mathrm{OPE}}(p,p^{\prime })=i\int
d^{4}xd^{4}ye^{ip^{\prime }y}e^{-ipx}\left\{ \mathrm{Tr}\left[ \gamma
_{5}S_{b}^{ja}(-x)\right. \right.  \notag \\
&&\left. \times \gamma _{5}\widetilde{S}_{c}^{ib}(y-x)\gamma _{\nu }%
\widetilde{S}_{b}^{ai}(x-y)\gamma _{\mu }S_{c}^{bj}(x)\right]  \notag \\
&&\left. -\mathrm{Tr}\left[ \gamma _{5}S_{b}^{ja}(-x)\gamma _{5}\widetilde{S}%
_{c}^{ib}(y-x)\gamma _{\nu }\widetilde{S}_{b}^{bi}(x-y)\gamma _{\mu
}S_{c}^{aj}(x)\right] \right\} .  \notag \\
&&
\end{eqnarray}

To derive the sum rule for $g_{3}(q^{2})$ we utilize invariant amplitudes
corresponding to the components proportional to $g_{\mu \nu }$ both in $\Pi
_{\mu \nu }^{3\mathrm{Phys}}(p,p^{\prime })$ and $\Pi _{\mu \nu }^{3\mathrm{%
OPE}}(p,p^{\prime })$. Numerical computations have been performed using the
following input parameters: For the parameters $M_{2}^{2}$ and$\
s_{0}^{\prime }$ in the $B_{c}^{\ast +}$ channel, we use
\begin{equation}
M_{2}^{2}\in \lbrack 6.5,7.5]~\mathrm{GeV}^{2},\ s_{0}^{\prime }\in \lbrack
50,51]~\mathrm{GeV}^{2}.
\end{equation}%
The decay constants $f_{B_{c}^{\ast }}$ and $f_{B_{c}}$ of the mesons $%
B_{c}^{\ast +}$ and $B_{c}^{-}$ have been chosen as $471~\mathrm{MeV}$ and $%
(371\pm 37)~\mathrm{MeV}$ \cite{Eichten:2019gig,Wang:2024fwc}, respectively.

The extrapolating function $\mathcal{G}_{3}(Q^{2})$ has the parameters: $%
\mathcal{G}_{3}^{0}=0.41~\mathrm{GeV}^{-1}$, $a_{3}^{1}=4.28$, and $%
a_{3}^{2}=-1.82$. For the strong coupling $g_{3}$, we get
\begin{equation}
g_{3}\equiv \mathcal{G}_{3}(-m_{B_{c}}^{2})=(1.3\pm 0.2)\times 10^{-1}\
\mathrm{GeV}^{-1}.
\end{equation}%
The partial width of the decay $T\rightarrow B_{c}^{\ast +}B_{c}^{-}$ is
equal to

\begin{equation}
\Gamma \left[ T\rightarrow B_{c}^{\ast +}B_{c}^{-}\right] =(21.6\pm 5.6)~%
\mathrm{MeV}.  \label{eq:DW3}
\end{equation}%
Omitting further details let us provide the width of the process $%
T\rightarrow B_{c}^{\ast -}B_{c}^{+}$ which is
\begin{equation}
\Gamma \left[ T\rightarrow B_{c}^{\ast -}B_{c}^{+}\right] =(15.6\pm 4.0)~%
\mathrm{MeV}.  \label{eq:DW4}
\end{equation}


\section{Decays generated by $b\overline{b}$ annihilation}

\label{sec:ScalarWidths2}


In this section of the article, we concentrate on decays of the tetraquark $%
T $ to the charmed meson pairs. These processes are triggered by the $b$ and
$\overline{b}$ annihilation to the light quarks $q\overline{q}$ followed by
creation of $D$ meson pairs with properly fixed charges and quantum numbers.
It is not difficult to see that production of $D^{0}\overline{D}^{\ast 0}$, $%
D^{\ast 0}\overline{D}^{0}$, $D^{\ast +}D^{-}$, and $D^{+}D^{\ast -}$ mesons
are among the allowed decay channels of the tetraquark $T$. In the framework
of the sum rule method these decays can be analyzed by employing the
three-point correlators in which the heavy quark vacuum matrix element $%
\langle \overline{b}b\rangle $ is replaced by the gluon condensate $\langle
\alpha _{s}G^{2}/\pi \rangle $.


\subsection{Decays $T\rightarrow D^{0}\overline{D}^{\ast 0}$ and $D^{\ast 0}%
\overline{D}^{0}$}


Here, we consider, in extended form, the process $T\rightarrow D^{0}%
\overline{D}^{\ast 0}$. In this case, the coupling $G_{1}$ describes the
strong interaction of particles at the vertex $TD^{0}\overline{D}^{\ast 0}$,
and can be obtained from the three-point correlation function%
\begin{eqnarray}
\widetilde{\Pi }_{\mu \nu }^{1}(p,p^{\prime }) &=&i^{2}\int
d^{4}xd^{4}ye^{ip^{\prime }y}e^{-ipx}\langle 0|\mathcal{T}\{J_{\nu }^{%
\overline{D}^{\ast 0}}(y)  \notag \\
&&\times J^{D^{0}}(0)J_{\mu }^{\dagger }(x)\}|0\rangle ,  \label{eq:CF1A}
\end{eqnarray}%
where the currents $J_{\nu }^{\overline{D}^{\ast 0}}(x)$ and $J^{D^{0}}(x)$
for the mesons $\overline{D}^{\ast 0}$ and $D^{0}$ are defined as
\begin{equation}
J^{D^{0}}(x)=\overline{u}_{j}(x)i\gamma _{5}c_{j}(x),\ J_{\nu }^{\overline{D}%
^{\ast 0}}(x)=\overline{c}_{i}(x)\gamma _{\nu }u_{i}(x).  \label{eq:CRB}
\end{equation}

\begin{figure}[h]
\includegraphics[width=8.5cm]{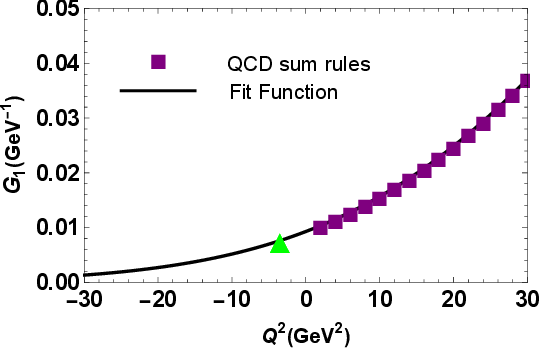}
\caption{The sum rule data and extrapolating function for the form factor $%
G_{1}(Q^{2})$. The triangle shows the point $Q^{2}=-m_{D^0}^{2}$. }
\label{fig:Fit1}
\end{figure}

In the context of SR method, we express the function $\widetilde{\Pi }_{\mu
\nu }^{1}(p,p^{\prime })$ using parameters of the particles $T$, $D^{0}$,
and $\overline{D}^{\ast 0}$
\begin{eqnarray}
&&\widetilde{\Pi }_{\mu \nu }^{1\mathrm{Phys}}(p,p^{\prime })=\frac{\langle
0|J_{\nu }^{\overline{D}^{\ast 0}}|\overline{D}^{\ast 0}(p^{\prime
},\varepsilon )\rangle }{p^{\prime 2}-m_{D^{\ast 0}}^{2}}\frac{\langle
0|J^{D^{0}}|D^{0}(q)\rangle }{q^{2}-m_{D^{0}}^{2}}  \notag  \label{eq:CF2} \\
&&\times \langle \overline{D}^{\ast 0}(p^{\prime },\varepsilon
)D^{0}(q)|T(p,\epsilon )\rangle \frac{\langle T(p,\epsilon )|J_{\mu
}^{\dagger }|0\rangle }{p^{2}-m^{2}}+\cdots ,  \notag \\
&&
\end{eqnarray}%
where $m_{D^{\ast 0}}=(2006.85\pm 0.05)~\mathrm{MeV}$ and $%
m_{D^{0}}=(1864.84\pm 0.05)~\mathrm{MeV}$ are the masses of the mesons $%
\overline{D}^{\ast 0}$ and $D^{0}$. Here, $\varepsilon _{\nu }=\varepsilon
_{\nu }(p^{\prime })$ is the polarization vector of $\overline{D}^{\ast 0}$.

To recast $\widetilde{\Pi }_{\mu \nu }^{1\mathrm{Phys}}(p,p^{\prime })$ into
its final form, we introduce the matrix elements
\begin{eqnarray}
&&\langle 0|J^{D^{0}}|D^{0}\rangle =\frac{f_{D}m_{D^{0}}^{2}}{m_{c}},  \notag
\\
&&\langle 0|J_{\nu }^{\overline{D}^{\ast 0}}|\overline{D}^{\ast 0}(p^{\prime
},\varepsilon )\rangle =f_{D^{\ast }}m_{D^{\ast 0}}\varepsilon _{\nu
}(p^{\prime }),  \label{eq:ME2B}
\end{eqnarray}%
with $f_{D}=(211.9\pm 1.1)~\mathrm{MeV}$ \cite{Rosner:2015wva} and $%
f_{D^{\ast }}=(252.2\pm 22.66)~\mathrm{MeV}$ being the decay constants of
the mesons $D^{0}$ and $\overline{D}^{\ast 0}$. The vertex $\langle
\overline{D}^{\ast 0}(p^{\prime },\varepsilon )D^{0}(q)|T(p,\epsilon
)\rangle $ is considered in the form
\begin{eqnarray}
&&\langle \overline{D}^{\ast 0}(p^{\prime },\varepsilon
)D^{0}(q)|T(p,\epsilon )\rangle =G_{1}(q^{2})\left[ p\cdot q\epsilon \cdot
\varepsilon ^{\ast }\right.  \notag \\
&&\left. -p\cdot \varepsilon ^{\ast }q\cdot \epsilon \right] .
\label{eq:ME3B}
\end{eqnarray}

The correlator $\widetilde{\Pi }_{\mu \nu }^{1\mathrm{Phys}}(p,p^{\prime })$
obtained by employing these matrix elements contains terms with the Lorentz
structures $g_{\mu \nu }$, $p_{\mu }p_{\nu },\ p_{\mu }^{\prime }p_{\nu
}^{\prime },$ $p_{\mu }^{\prime }p_{\nu }$ and $p_{\mu }p_{\nu }^{\prime }$.
The SR for the form factor $G_{1}(q^{2})$ is derived using invariant
amplitude $\widetilde{\Pi }^{1\mathrm{Phys}}(p^{2},p^{\prime 2},q^{2})$ that
corresponds to the structure $g_{\mu \nu }$.

The QCD side of the SR for $G_{1}(q^{2})$ is given by the formula
\begin{eqnarray}
&&\widetilde{\Pi }_{\mu \nu }^{1\mathrm{OPE}}(p,p^{\prime })=-\frac{2i}{3}%
\int d^{4}xd^{4}ye^{ip^{\prime }y}e^{-ipx}\langle \overline{b}b\rangle
\notag \\
&&\times \mathrm{Tr}\left[ \gamma _{\nu }S_{u}^{ij}(y)\gamma
_{5}{}S_{c}^{jb}(-x)\gamma _{5}\gamma _{\mu }S_{c}^{bi}(x-y)\right] ,
\label{eq:QCDsideA}
\end{eqnarray}%
where $S_{u}(x)$ is the propagator of $u$ quark \cite{Agaev:2020zad}.

For further studies we apply the relation between heavy-quark and gluon
condensates derived in the context of the QCD sum rule method \cite%
{Shifman:1978bx}
\begin{equation}
\langle \overline{b}b\rangle =-\frac{1}{12m_{b}}\langle \frac{\alpha
_{s}G^{2}}{\pi }\rangle .  \label{eq:Conden}
\end{equation}%
This equality is valid provided the higher orders in $1/m_{b}$ are small and
can be neglected. This substitution leads to the expression
\begin{eqnarray}
&&\widetilde{\Pi }_{\mu \nu }^{1\mathrm{OPE}}(p,p^{\prime })=\frac{i}{18m_{b}%
}\langle \frac{\alpha _{s}G^{2}}{\pi }\rangle \int d^{4}xd^{4}ye^{ip^{\prime
}y}e^{-ipx}  \notag \\
&&\times \mathrm{Tr}\left[ \gamma _{\nu }S_{u}^{ij}(y)\gamma
_{5}{}S_{c}^{jb}(-x)\gamma _{5}\gamma _{\mu }S_{c}^{bi}(x-y)\right] .
\label{eq:CF6A}
\end{eqnarray}%
In what follows, we denote by $\widetilde{\Pi }^{1\mathrm{OPE}%
}(p^{2},p^{\prime 2},q^{2})$ the invariant amplitude which corresponds to
the structure $g_{\mu \nu }$.

The SR for the coupling $G_{1}(q^{2})$ reads%
\begin{eqnarray}
&&G_{1}(q^{2})=\frac{2m_{c}}{\Lambda f_{D}m_{D^{0}}^{2}f_{D^{\ast
}}m_{D^{\ast 0}}}\frac{q^{2}-m_{D^{0}}^{2}}{m^{2}-m_{D^{\ast 0}}^{2}+q^{2}}
\notag \\
&&\times e^{m^{2}/M_{1}^{2}}e^{m_{D^{\ast 0}}^{2}/M_{2}^{2}}\widetilde{\Pi }%
^{1}(\mathbf{M}^{2},\mathbf{s}_{0},q^{2}),
\end{eqnarray}%
where $\widetilde{\Pi }^{1}(\mathbf{M}^{2},\mathbf{s}_{0},q^{2})$ is the
amplitude $\widetilde{\Pi }^{1\mathrm{OPE}}(p^{2},p^{\prime 2},q^{2})$
undergone to Borel transformations and continuum subtractions.

To extract $G_{1}(q^{2})$ from this SR, we carry out standard manipulations,
and skip further details: In the $\overline{D}^{\ast 0}$ meson channel, we
have used the parameters
\begin{equation}
M_{2}^{2}\in \lbrack 2,3]~\mathrm{GeV}^{2},\ s_{0}^{\prime }\in \lbrack
5.7,5.8]~\mathrm{GeV}^{2}.  \label{eq:Wind2}
\end{equation}%
The coupling $G_{1}$ has been evaluated by employing SR data for $%
Q^{2}=2-30\ \mathrm{GeV}^{2}$ and extrapolating function with parameters $%
\widetilde{\mathcal{G}}_{1}^{0}=0.0117~\mathrm{GeV}^{-1}$, $\widetilde{a}%
_{1}^{1}=8.95$, and $\widetilde{a}_{1}^{2}=-7.99$. The SR data and fit
function $\widetilde{\mathcal{G}}_{1}(Q^{2})$ are plotted in Fig.\ \ref%
{fig:Fit1}. The coupling $G_{1}$ is evaluated at the mass shell $%
q^{2}=m_{D^{0}}^{2}$ and amounts to
\begin{equation}
G_{1}\equiv \widetilde{\mathcal{G}}_{1}(-m_{D^{0}}^{2})=(9.60\pm 2.15)\times
10^{-3}\ \mathrm{GeV}^{-1}.  \label{eq:G1}
\end{equation}%
The width of the decay $T\rightarrow D^{0}\overline{D}^{\ast 0}$ is
\begin{equation}
\Gamma \left[ T\rightarrow D^{0}\overline{D}^{\ast 0}\right] =(11.5\pm 3.6)~%
\mathrm{MeV}.  \label{eq:DW5}
\end{equation}

Parameters of the process $T\rightarrow D^{\ast 0}\overline{D}^{0}$ can be
computed by this way as well. In this case, the coupling $G_{2}$ at the
vertex $TD^{\ast 0}\overline{D}^{0}$ is
\begin{equation}
G_{2}\equiv \widetilde{\mathcal{G}}_{1}(-m_{D^{0}}^{2})=(6.23\pm 1.31)\times
10^{-3}\ \mathrm{GeV}^{-1}.
\end{equation}%
Then the width of this decay becomes equal to
\begin{equation}
\Gamma \left[ T\rightarrow D^{\ast 0}\overline{D}^{0}\right] =(4.8\pm 1.5)~%
\mathrm{MeV}.  \label{eq:DW6}
\end{equation}


\subsection{Processes $T\rightarrow D^{\ast +}D^{-}$ and $D^{+}D^{\ast -}$}


The modes $T\rightarrow D^{\ast +}D^{-}$ and $D^{+}D^{\ast -}$ are explored
in accordance with the scheme explained above. Let us consider the process $%
T\rightarrow D^{\ast +}D^{-}$. The strong coupling $G_{3}$ at the vertex $%
TD^{\ast +}D^{-}$ can be extracted from the correlation function
\begin{eqnarray}
\widetilde{\Pi }_{\mu \nu }^{3}(p,p^{\prime }) &=&i^{2}\int
d^{4}xd^{4}ye^{ip^{\prime }y}e^{-ipx}\langle 0|\mathcal{T}\{J_{\nu
}^{D^{\ast +}}(y)  \notag \\
&&\times J^{D^{-}}(0)J_{\mu }^{\dagger }(x)\}|0\rangle .
\end{eqnarray}%
Here, $J_{\nu }^{D^{\ast +}}(x)$ and $J^{D^{-}}(x)$ are the interpolating
currents of the $D^{\ast +}$ and $D^{-}$ mesons
\begin{equation}
J_{\nu }^{D^{\ast +}}(x)=\overline{d}_{i}(x)\gamma _{\nu }c_{i}(x),\
J^{D^{-}}(x)=\overline{c}_{j}(x)i\gamma _{5}d_{j}(x).
\end{equation}%
The matrix elements of these particles required to calculate $\widetilde{\Pi
}_{\mu \nu }^{3\mathrm{Phys}}(p,p^{\prime })$ are%
\begin{eqnarray}
&&\langle 0J_{\nu }^{D^{\ast +}}|D^{\ast +}(p^{\prime },\varepsilon )\rangle
=f_{D^{\ast }}m_{D^{\ast }}\varepsilon _{\nu }(p^{\prime }),  \notag \\
&&\langle 0|J^{D^{-}}|D^{-}(q)\rangle =\frac{f_{D}m_{D}^{2}}{m_{c}}.
\label{eq:ME5}
\end{eqnarray}%
In Eq.\ (\ref{eq:ME5}) $m_{D^{\ast }}=(2010.26\pm 0.05)~\mathrm{MeV}$ and $%
m_{D}=(1869.5\pm 0.4)~\mathrm{MeV}$ are the masses of the mesons $D^{\ast
\pm }$ and $D^{\pm }$, respectively. The matrix element $\langle D^{\ast
+}(p^{\prime },\varepsilon )D^{-}(q)|T(p,\epsilon )\rangle $ of
axial-vector, vector and pseudoscalar particles has been defined above.
Therefore, we skip these questions and write down QCD side of the SR
\begin{eqnarray}
&&\widetilde{\Pi }_{\mu \nu }^{3\mathrm{OPE}}(p,p^{\prime })=\frac{i}{18m_{b}%
}\langle \frac{\alpha _{s}G^{2}}{\pi }\rangle \int d^{4}xd^{4}ye^{ip^{\prime
}y}e^{-ipx}  \notag \\
&&\times \mathrm{Tr}\left[ \gamma _{\nu }{}S_{c}^{ib}(y-x)\gamma _{5}\gamma
_{\mu }S_{c}^{bj}(x)\gamma _{5}S_{d}^{ji}(-y)\right] .
\end{eqnarray}

As usual, we utilize invariant amplitudes which correspond to structures $%
g_{\mu \nu }$. In numerical calculations the Borel and continuum subtraction
parameters in the $D^{\ast +}$ channel are fixed as in Eq.\ (\ref{eq:Wind2}%
). The coupling $G_{3}$ is
\begin{equation}
G_{3}\equiv \widetilde{\mathcal{G}}_{3}(-m_{D}^{2})=(6.23\pm 1.31)\times
10^{-3}\ \mathrm{GeV}^{-1}.
\end{equation}%
For the partial width of the mode $T\rightarrow D^{\ast +}D^{-}$, we get
\begin{equation}
\Gamma \left[ T\rightarrow D^{\ast +}D^{-}\right] =(4.8\pm 1.5)~\mathrm{MeV}.
\label{eq:DW7}
\end{equation}%
The process $T\rightarrow D^{+}D^{\ast -}$ can be explored in the same
manner. The coupling $G_{4}$ responsible for strong interaction of the
particles at the vertex $TD^{+}D^{\ast -}$ is%
\begin{equation}
G_{4}\equiv \widetilde{\mathcal{G}}_{4}(-m_{D}^{2})=(9.60\pm 2.15)\times
10^{-3}\ \mathrm{GeV}^{-1}.
\end{equation}%
For the width of this decay, we find
\begin{equation}
\Gamma \left[ T\rightarrow D^{+}D^{\ast -}\right] =(11.5\pm 3.6)~\mathrm{MeV}%
.  \label{eq:DW8}
\end{equation}

As is seen, the partial widths of the processes $T\rightarrow D^{+}D^{\ast
-} $ and $D^{0}\overline{D}^{\ast 0}$, as well as of the decays $%
T\rightarrow D^{\ast +}D^{-}$ and $D^{\ast 0}\overline{D}^{0}$ are equal to
each other. This is not surprising output because replacements $d\rightarrow
u$ in the mesons $D^{+}D^{\ast -}$ transform them to the $D^{0}\overline{D}%
^{\ast 0}$ particles. Similarly, the final states $D^{\ast +}D^{-}$ and $%
D^{\ast 0}\overline{D}^{0}$ are connected by the $u\leftrightarrow d$
transforms. Because, we work in the approximation $m_{u}=m_{d}$ the
parameters of these decays coincide with each other: Effects of differences
in the masses of the mesons on final results are negligible. This fact is
used in the next part of this article, when we consider the decays triggered
by the annihilation of $c$ and $\overline{c}$ quarks in $T$.


\section{Decays due to $c\overline{c}$ annihilation}

\label{sec:ScalarWidths3}


Here, we explore the processes $T\rightarrow B^{\ast +}B^{-}$, $B^{+}B^{\ast
-}$, $B^{\ast 0}\overline{B}^{0}$, and $B^{0}\overline{B}^{\ast 0}$ which
become possible due to annihilation of the constituent quarks $c\overline{c}$
inside of the tetraquark $T$. These decays are considered in accordance with
a technique described in the previous section difference being in the vacuum
expectation value $\langle \overline{c}c\rangle $ of the $c$ quarks, which
takes the form
\begin{equation}
\langle \overline{c}c\rangle =-\frac{1}{12m_{c}}\langle \frac{\alpha
_{s}G^{2}}{\pi }\rangle .
\end{equation}%
It is not difficult to see that in the approximation $m_{u}=m_{d}$ the
widths of the processes $T\rightarrow B^{\ast +}B^{-}$ and $T\rightarrow
B^{\ast 0}\overline{B}^{0}$ are equal to each other. The same conclusion is
valid for the widths of the decays $T\rightarrow B^{+}B^{\ast -}$ and $%
T\rightarrow B^{0}\overline{B}^{\ast 0}$ as well. Therefore, we explore and
compute explicitly the parameters of the reactions $T\rightarrow B^{\ast
+}B^{-}$ and $B^{+}B^{\ast -}$, and by utilizing aforementioned relations
estimate parameters of the remaining two decays.


\subsection{$T\rightarrow B^{\ast +}B^{-}$ and $T\rightarrow B^{\ast 0}%
\overline{B}^{0}$}


We start from investigation of the decay $T\rightarrow B^{\ast +}B^{-}$. The
SR for the form factor $F_{1}(q^{2})$ at the vertex $TB^{\ast +}B^{-}$ can
be derived from analysis of the correlation function
\begin{eqnarray}
\widehat{\Pi }_{\mu \nu }^{1}(p,p^{\prime }) &=&i^{2}\int
d^{4}xd^{4}ye^{ip^{\prime }y}e^{-ipx}\langle 0|\mathcal{T}\{J_{\nu
}^{B^{\ast +}}(y)  \notag \\
&&\times J^{B^{-}}(0)J_{\mu }^{\dagger }(x)\}|0\rangle ,
\end{eqnarray}%
where $J_{\nu }^{B^{\ast +}}(x)$ and $J^{B^{-}}(x)$ are the interpolating
currents of the $B^{\ast +}$ and $B^{-}$ mesons
\begin{equation}
J_{\nu }^{B^{\ast +}}(x)=\overline{b}_{i}(x)\gamma _{\nu }u_{i}(x),\
J^{B^{-}}(x)=\overline{u}_{j}(x)i\gamma _{5}b_{j}(x).
\end{equation}

The physical side of the SRs for the form factor $F_{1}(q^{2})$ is given by
the formula
\begin{eqnarray}
&&\widehat{\Pi }_{\mu \nu }^{1\mathrm{Phys}}(p,p^{\prime })=\frac{\langle
0|J_{\nu }^{B^{\ast +}}|B^{\ast +}(p^{\prime },\varepsilon )\rangle }{%
p^{\prime 2}-m_{B^{\ast }}^{2}}\frac{\langle 0|J^{B^{-}}|B^{-}(q)\rangle }{%
q^{2}-m_{B}^{2}}  \notag \\
&&\times \langle B^{\ast +}(p^{\prime },\varepsilon )B^{-}(q)|T(p,\epsilon
)\rangle \frac{\langle T(p,\epsilon )|J_{\mu }^{\dagger }|0\rangle }{%
p^{2}-m^{2}}+\cdots ,  \notag \\
&&
\end{eqnarray}%
where the matrix elements necessary to calculate $\widehat{\Pi }_{\mu \nu
}^{1\mathrm{Phys}}(p,p^{\prime })$ are%
\begin{eqnarray}
&&\langle 0|J_{\nu }^{B^{\ast +}}|B^{\ast +}(p^{\prime },\varepsilon
)\rangle =f_{B^{\ast }}m_{B^{\ast }}\varepsilon _{\nu }(p^{\prime }),  \notag
\\
&&\langle 0|J^{B^{-}}|B^{-}(q)\rangle =\frac{f_{B}m_{B}^{2}}{m_{b}}.
\end{eqnarray}%
Here $m_{B}=(5279.42\pm 0.08)~\mathrm{MeV}$ and $m_{B^{\ast }}=(5324.75\pm
0.20)~\mathrm{MeV}$ are the masses of the $B^{-}$ and $B^{\ast +}$ mesons
\cite{PDG:2022}. For their decay constants $f_{B}=(206\pm 7)~\mathrm{MeV}$
and $f_{B^{\ast }}=(210\pm 6)~\mathrm{MeV}$ we use information from Ref.\
\cite{Narison:2015nxh}. The matrix element $\langle B^{\ast +}(p^{\prime
},\varepsilon )B^{-}(q)|T(p,\epsilon )\rangle $ has been introduced in
previous parts of the paper. Then, the function $\widehat{\Pi }_{\mu \nu
}^{1\mathrm{Phys}}(p,p^{\prime })$ takes the form
\begin{eqnarray}
&&\widehat{\Pi }_{\mu \nu }^{1\mathrm{Phys}}(p,p^{\prime })=\frac{%
F_{1}(q^{2})\Lambda f_{B^{\ast }}m_{B^{\ast }}f_{B}m_{B}^{2}}{m_{b}\left(
p^{2}-m^{2}\right) \left( p^{\prime 2}-m_{B^{\ast }}^{2}\right) }  \notag \\
&&\times \frac{1}{(q^{2}-m_{B}^{2})}\left[ \frac{m^{2}-m_{B^{\ast
}}^{2}+q^{2}}{2}g_{\mu \nu }-p_{\mu }p_{\nu }+\cdots \right] .  \notag \\
&&
\end{eqnarray}

The QCD side are determined by the expression
\begin{eqnarray}
&&\widehat{\Pi }_{\mu \nu }^{1\mathrm{OPE}}(p,p^{\prime })=\frac{i}{18m_{c}}%
\langle \frac{\alpha _{s}G^{2}}{\pi }\rangle \int d^{4}xd^{4}ye^{ip^{\prime
}y}e^{-ipx}  \notag \\
&&\times \mathrm{Tr}\left[ \gamma _{\nu }{}S_{u}^{ij}(y)\gamma
_{5}S_{b}^{ja}(-x)\gamma _{5}\gamma _{\mu }S_{b}^{ai}(x-y)\right] .
\end{eqnarray}%
The SR for the form factor $F_{1}(q^{2})$ is
\begin{eqnarray}
&&F_{1}(q^{2})=\frac{2m_{b}}{\Lambda f_{B^{\ast }}m_{B^{\ast }}f_{B}m_{B}^{2}%
}\frac{q^{2}-m_{B}^{2}}{m^{2}-m_{B^{\ast }}^{2}+q^{2}}  \notag \\
&&\times e^{m^{2}/M_{1}^{2}}e^{m_{B^{\ast }}^{2}/M_{2}^{2}}\widehat{\Pi }%
^{1}(\mathbf{M}^{2},\mathbf{s}_{0},q^{2}),
\end{eqnarray}

\begin{figure}[h]
\includegraphics[width=8.5cm]{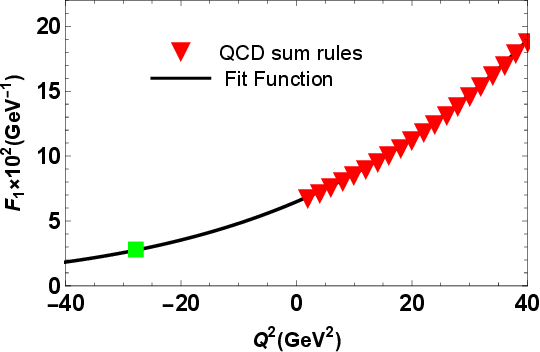}
\caption{The QCD data and fitting function for the form factor $F_{1}(Q^{2})$%
. The square locates at the point $Q^{2}=-m_{B}^{2}$. }
\label{fig:Fit2}
\end{figure}

To carry out the numerical computations, we utilize amplitudes that
correspond to structures $g_{\mu \nu }$. The Borel and continuum subtraction
parameters in the $T$ channel are fixed as in Eq.\ (\ref{eq:Wind1}). The
parameters $(M_{2}^{2},s_{0}^{\prime })$ in the channel of the $B^{\ast +}$
meson are chosen within limits
\begin{equation}
M_{2}^{2}\in \lbrack 5.5,6.5]~\mathrm{GeV}^{2},\ s_{0}^{\prime }\in \lbrack
34,35]~\mathrm{GeV}^{2}.  \label{eq:Wind4}
\end{equation}%
The form factor $F_{1}(Q^{2})$ is calculated at $Q^{2}=2-40~\mathrm{GeV}%
^{2}\ $and extrapolated by the function $\widehat{\mathcal{G}}_{1}(Q^{2})$
with parameters $\widehat{\mathcal{G}}_{1}^{0}=0.065~\mathrm{GeV}^{-1}$, $%
\widehat{a}_{1}^{1}=4.72$, and $\widehat{a}_{1}^{2}=-1.57$. The
extrapolating function $\widehat{\mathcal{G}}_{1}(Q^{2})$ and QCD data are
shown in Fig.\ \ref{fig:Fit2}. At the mass shell of the $B^{-}$ meson $%
q^{2}=m_{B}^{2}$ this function leads to the following strong coupling $F_{1}$
\begin{equation}
F_{1}\equiv \widehat{\mathcal{G}}_{1}(-m_{B}^{2})=(2.8\pm 0.6)\times
10^{-2}\ \mathrm{GeV}^{-1}.
\end{equation}%
For the partial width of the mode $T\rightarrow B^{\ast +}B^{-}$, we get
\begin{equation}
\Gamma \left[ T\rightarrow B^{\ast +}B^{-}\right] =(6.3\pm 1.9)~\mathrm{MeV}.
\label{eq:DW9}
\end{equation}%
The width of the second process $T\rightarrow B^{\ast 0}\overline{B}^{0}$
with a high accuracy is equal to $\Gamma \left[ T\rightarrow B^{\ast +}B^{-}%
\right] $.

\subsection{$T\rightarrow B^{+}B^{\ast -}$ and $B^{0}\overline{B}^{\ast 0}$}


The decays $T\rightarrow B^{+}B^{\ast -}$ and $B^{0}\overline{B}^{\ast 0}$
are studied in the same way. In the case of the process $T\rightarrow
B^{+}B^{\ast -}$ the correlation function $\widehat{\Pi }_{\mu \nu }^{2%
\mathrm{OPE}}(p,p^{\prime })$ is given by the formula%
\begin{eqnarray}
&&\widehat{\Pi }_{\mu \nu }^{2\mathrm{OPE}}(p,p^{\prime })=\frac{i}{18m_{c}}%
\langle \frac{\alpha _{s}G^{2}}{\pi }\rangle \int d^{4}xd^{4}ye^{ip^{\prime
}y}e^{-ipx}  \notag \\
&&\times \mathrm{Tr}\left[ \gamma _{\nu }{}S_{b}^{ia}(y-x)\gamma _{5}\gamma
_{\mu }S_{b}^{aj}(x)\gamma _{5}S_{u}^{ji}(-y)\right] .
\end{eqnarray}%
The QCD data are calculated using the Borel and continuum subtraction
parameters presented above. Therefore, below we provide final results for
the fitting function $\widehat{\mathcal{G}}_{2}(Q^{2})$ which are determined
by the parameters $\widehat{\mathcal{G}}_{2}^{0}=0.066~\mathrm{GeV}^{-1}$, $%
\widehat{a}_{2}^{1}=5.06$, and $\widehat{a}_{2}^{2}=-1.94$. At the mass
shell $q^{2}=m_{B}^{2}$ it equals to
\begin{equation}
F_{2}\equiv \widehat{\mathcal{G}}_{2}(-m_{B}^{2})=(2.6\pm 0.5)\times
10^{-2}\ \mathrm{GeV}^{-1}.
\end{equation}
which is the strong coupling $F_{2}$ at the vertex $TB^{+}B^{\ast -}$. The
width of the decay $T\rightarrow B^{+}B^{\ast -}$ amounts to
\begin{equation}
\Gamma \left[ T\rightarrow B^{+}B^{\ast -}\right] =(5.6\pm 1.6)~\mathrm{MeV}.
\label{eq:DW10}
\end{equation}
The partial width of the second process $T\rightarrow B^{0}\overline{B}%
^{\ast 0}$ is also determined by Eq.\ (\ref{eq:DW10}).

Computations performed in the current article permit us to estimate the full
width of the axial-vector tetraquark $T$ with content $bc\overline{b}%
\overline{c}$. As a result, we obtain%
\begin{equation}
\Gamma \left[ T\right] =(140\pm 13)~\mathrm{MeV},
\end{equation}%
which characterizes it as a relatively broad resonance.


\section{Concluding notes}

\label{sec:Conc}


In present article, we have calculated the mass and full width of the
axial-vector tetraquark $T$. Analyses have been performed in the framework
of QCD sum rule method. To evaluate the mass of $T$, we have applied the
two-point SR method, whereas its decays have been studied by invoking the
three-point SR approach. The axial-vector tetraquark $T$ \ together with the
scalar diquark-antidiquark state $T_{\mathrm{bc\overline{b}\overline{c}}}$
forms a family of fully heavy exotic mesons with fixed content but different
spin-parities. The particle $T_{\mathrm{bc\overline{b}\overline{c}}}$ was
considered in our paper \cite{Agaev:2024wvp} and has parameters which are
close to ones extracted in this article.

The mass $m$ of the tetraquark $T$ was calculated in different publications:
We should note a wide diversity of predictions obtained in these papers. Our
result is smaller than the ones extracted using different models and
methods. Only in a few articles the authors found $m$ below $m=(12715\pm 90)~%
\mathrm{MeV}$. One of main reasons to investigate fully heavy
diquark-antidiquark states is a desire to find four-quark mesons which would
be stable against strong decays. Conclusions about the stability of some of
the considered structures against the strong dissociations to two-mesons
were made in those articles. But it has to be emphasized that exotic mesons $%
bc\overline{b}\overline{c}$ due to features of their contents are, in any
case, strong-interaction unstable particles, because $b\overline{b}$ and $c%
\overline{c}$ annihilations to light quark pairs generate their decays to
the ordinary mesons.

Our results characterize $T$ as a relatively broad four-quark structure,
which can decay to two-meson final states through both dissociation and $b%
\overline{b}$, $c\overline{c}$ annihilation mechanisms. We have calculated
widths of four such fall-apart processes, which are dominant decay channels
of $T$. We have evaluated also widths of modes triggered by $b\overline{b}$
and $c\overline{c}$ annihilations inside $T$. Contributions of these
processes is considerable and amount to approximately $23\%$ and $17\%$ of
the $T$ tetraquark's full width, respectively.

Investigations of fully heavy tetraquarks are important for ongoing and
planning experiments. Calculations of their parameters are necessary for
reliable interpretation of collected experimental data. Existing
publications are concentrated mainly on masses of such four-quark
structures: Decays of these states, including $bc\overline{b}\overline{c}$
ones, were not attracted strong interest of researchers. But without this
information all analyses remain incomplete and do not allow one to choose
credible predictions and methods. Evidently new studies are required to
clarify remaining problems.

\end{document}